\documentclass[twocolumn,showpacs,prb]{revtex4}
\usepackage{amssymb}

\usepackage{graphicx}
\usepackage{epsfig}
\usepackage[T1]{fontenc}

\newcommand{\beq}{\begin{equation}}
\newcommand{\eeq}{\end{equation}}
\newcommand{\bea}{\begin{eqnarray}}
\newcommand{\eea}{\end{eqnarray}}

\newcommand{\veps}{\varepsilon}
\newcommand{\al}{\alpha}
\newcommand{\s}{\sigma}

\newcommand{\ga}{\gamma}

\newcommand{\ua}{\uparrow}
\newcommand{\da}{\downarrow}

\newcommand{\tht}{\theta}
\newcommand{\pbf}{}

\begin{document}

\title{Kondo effect with non collinear polarized leads: \\
a numerical renormalization group analysis}
\author{P. Simon$^1$, P.~S. Cornaglia$^{2}$, D. Feinberg$^3$, and C.~A. Balseiro$^4$}

\affiliation{$^1$Laboratoire de Physique et Mod\'elisation des Milieux 
Condens\'es, CNRS and Universit\'e Joseph Fourier, BP 166, 38042 Grenoble,
France}

\affiliation{$^2$Centre de Physique Théorique, École Polytechnique, CNRS, 91128
Palaiseau cedex, France}

\affiliation{$^3$Laboratoire d'Etudes des Propri\'et\'es Electroniques des
Solides, CNRS, associated to Universit\'e Joseph Fourier, BP 166, 38042
Grenoble, France}

\affiliation{$^4$Instituto Balseiro and Centro At\'{o}mico Bariloche, Comisi\'{o}n 
Nacional de Energ\'{\i}a At\'{o}mica, 8400 Bariloche, Argentina}

\date{\today}

\begin{abstract}
The Kondo effect in quantum dots attached to ferromagnetic leads with
general polarization directions is studied combining poor man scaling and Wilson's
numerical renormalization group methods. We show that polarized
electrodes will lead in general to a splitting of the Kondo resonance in the quantum dot
density of states 
except for a small range of angles close to the 
antiparallel case. We also show that an external magnetic field
is able to compensate this splitting and restore the unitary limit.
Finally, we study the electronic transport through the device in various limiting cases.
\end{abstract}

\maketitle



\section{Introduction} 
One of the most remarkable achievements of recent
progress in nanoelectronics has been the observation of the Kondo effect in
a single semiconductor quantum dot (QD).\cite{dot,Cronenwett} Since these
early experiments, the Kondo effect has also been observed in carbon
nanotube QDs,\cite{cobden,bachtold,jarillo} and molecular transistors\cite{molecular} 
connected to normal metallic leads. The success of these experiments has
opened the way to a systematic analysis, both theoretically and
experimentally, of the effects of the leads correlations on the Kondo effect.
Of particular interest are the experimental realizations of the Kondo effect
connected to superconducting leads\cite{supra} or to ferromagnetic
electrodes.\cite{ralph}  The latter is particularly important since it is
connected to the new emerging field called ``spintronics''.\cite{loss} One
goal of spintronics is to provide devices where the transport properties are
controlled by magnetic degrees of freedom. In this sense, a  Kondo quantum
dot connected to ferromagnetic electrodes may be regarded as an interesting
setup to study interaction effects in spintronics devices.

From the theoretical point of view, it has been predicted, despite some
early controversial results,\cite{sergueev,zhang,bulka,lopez1} that
ferromagnetic electrodes with parallel polarization produce a splitting of
the Kondo zero bias anomaly\cite{martinek1,martinek2,lopez2} whereas 
polarized electrodes with antiparallel polarization generally do not.\cite{martinek1,lopez2,utsumi} Some of these predictions have been checked
experimentally by contacting C$_{60}$ molecules with ferromagnetic nickel
electrodes. \cite{ralph} The underlying mechanism responsible of the
splitting of the Kondo resonance is quite clear: virtual quantum charge
fluctuations between the dot and the leads become spin dependent. As a
result, an effective exchange magnetic field appears in the dot which
generally lifts the spin degeneracy on it. For antiparallel
configurations, the contributions of each electrode to the effective field 
 may eventually cancel each other for symmetric couplings.

Most of the previous works have focused quite surprisingly only on parallel and
antiparallel polarizations. Sergueev et al. \cite{sergueev}
considered non collinear polarizations but did not take into account the
spin splitting of the dot energy level which naturally gives rise to
spurious results. In a recent preprint, Swirkowicz \textit{et al.}\cite{barnas}
analyze the non collinear situation using the equation of motion method
supplemented by an effective exchange field that has to be determined
self-consistently. Such method gives qualitative results for temperature $
T\sim T_K$, where $T_K$ is the Kondo temperature. The main result is that the splitting of
the zero bias Kondo anomaly decreases monotonically when increasing the
angle $2\theta$ between the two lead magnetizations. In this paper, we go
beyond the equation of motion method by combining a poor man scaling approach and
the numerical renormalization group (NRG) approach. The poor man scaling
provides us with an analytical expression for the Kondo temperature as a function
of the lead polarization $p$ and $\theta$ [see Eq.(\ref{tk})] while NRG
allows us to compute dot spectral densities and the
near equilibrium transport properties like the conductance.
The plan of the paper is the following: In Sec. \ref{sec:model} we describe
 our model Hamiltonian. In Sec. \ref{sec:rg}, we use poor man scaling to 
estimate the Kondo temperature. In Sec. \ref{sec:nrg}, we present the NRG 
results for the dot density of states and for the conductance. Finally, Sec. \ref{sec:conclusion}
contains a brief summary of the results. 

\section{Model Hamiltonian} \label{sec:model}
In order to describe this system we use the following general Hamiltonian 
\beq\label{eq:h}
H=H_{leads}+H_{tun}+H_{dot},
\eeq
 where $H_{leads}$ denotes the Hamiltonian of the conduction electrons in the leads 
and is defined by 
\beq
H_{\mathrm{leads}}=\sum\limits_{k,\gamma ,\pm }\varepsilon _{k\pm }c_{k,\gamma ,\pm
}^{\dag }c_{k,\gamma ,\pm }.
\eeq
In this equation $c_{k,\gamma ,\pm }^{\dag }$ creates an
electron with energy $\varepsilon _{k\pm }$ in lead $\gamma =L,R$ with spin
along $\pm \vec{n}_{\ga}$ where $\vec{n}_{\ga}$ is a unitary vector parallel
to the magnetization moment in lead $\gamma $. The angle between $\vec n_L$ and $\vec n_R$ 
is $2\theta$. Since the leads are assumed
to be polarized, there is a strong spin asymmetry in the density of states 
$\rho _{\gamma \pm }(\omega )$. 
The tunneling junctions between the leads and
the dot may be described by a standard tunneling Hamiltonian
\beq
H_{tun}=\sum\limits_{k,\gamma ,\s=\ua,\da
}(t_{\gamma }c_{k,\gamma ,\s}^{\dag }d_{\s}+H.c.),
\eeq 
where $d_{\s}$ destroys an
electron in the dot with spin $\s=\ua,\da$. Here the possible values for the dot spin are quantized along the axis $\vec{e}
_{z}=(\vec{n}_{L}+\vec{n}_{R})/|\vec{n}_{L}+\vec{n}_{R}|$. The operator $c_{k,\gamma ,\s}^{}$ is a linear combination of
the operators $c_{k,\gamma ,s=\pm}^{}$ and reads:
\beq
\left(
\begin{array}{c}
 c_{\al,k,\ua} \\  c_{\al,k,\da} \end{array}\right)=
\left( \begin{array}{cc}
\cos(\frac{\theta_\al}{2})&-\sin(\frac{\theta_\al}{2}) \\ \sin(\frac{\theta_\al}{2}) &\cos(\frac{\theta_\al}{2}) \end{array}\right)
\left(\begin{array}{c}
 c_{\al,k,+} \\  c_{\al,k,-} \end{array}\right),
\eeq
where $\theta_\al=\pm \theta$ here.
$H_{dot}$ is the usual Anderson
Hamiltonian and reads 
\begin{equation}\label{eq:anderson}
H_{dot}=\sum_{\sigma =\uparrow ,\downarrow }\varepsilon _{d,\sigma
}d_{s}^{\dag }d_{s}+Ud_{\ua}^{\dag }d_{\ua}d_{\da}^{\dag }d_{\da}-g\mu
_{B}BS^{z},
\end{equation}
where $S^{z}=(d_{\ua}^{\dag }d_{\ua}-d_{\da}^{\dag }d_{\da})$. The
last term of Eq. (\ref{eq:anderson}) 
describes the Zeeman energy splitting of the dot. In what follows
we mainly consider a symmetric situation in which $t_{\gamma }=t$. We denote
the total tunneling strength by $\Gamma =\Gamma _{L}+\Gamma _{R}$ with $
\Gamma _{L}=\pi t^{2}(\rho _{+}+\rho _{-})$. The effective magnetic fields
generated by the ferromagnetic electrodes is then along the axis $\vec{e}
_{z}=(\vec{n}_{L}+\vec{n}_{R})/|\vec{n}_{L}+\vec{n}_{R}|$. We next perform a
poor man scaling analysis.

\section{Poor man scaling analysis}\label{sec:rg} 
The analysis extends in a straightforward manner the
one developed in Ref. [\onlinecite{martinek1}]. In this part we neglect the energy
dependence in the density of states and also suppose $\rho _{\gamma \pm
}\approx \rho _{\pm }$. In the ferromagnetic leads, the ratio $p=(\rho
_{+}-\rho _{-})/(\rho _{+}+\rho _{-})$ is in general different from zero and
as shown bellow it is one of the relevant parameters to describe the effect
of the spin polarization. For an energy dependent density of states, 
the situation is more complex as will be shown in Sec. \ref{sec:nrg}. 
One performs a two stage renormalization group (RG)
analysis by reducing the cutoff $\Lambda $ from $D$ the bandwidth to $
\bar{\varepsilon}_{d}=\varepsilon _{d\uparrow }+\varepsilon _{d\downarrow }$
which renormalizes the parameters of the Anderson model. In general, $
\varepsilon _{\ua}$ and $\varepsilon _{\da}$ renormalize differently due to
the spin dependent density of states (DOS) in the leads.\cite{martinek1} The spin degeneracy of
the Anderson energy levels is therefore lifted  with an effective Zeeman
splitting 
\beq\delta \varepsilon _{d}=\varepsilon _{d\uparrow }-\varepsilon
_{d\downarrow }\sim p\Gamma \cos (\theta )\ln (D_{}/\Lambda ).
\eeq
 This
splitting prevents from reaching  of the strong coupling regime. Nonetheless, an external magnetic field 
$B$ can be applied locally to compensate this internal field and restore the
degeneracy between $\varepsilon _{\ua}$ and $\varepsilon _{\da}$. One can
then perform a Schrieffer-Wolff transformation. The Kondo Hamiltonian so
obtained is similar to the usual Kondo Hamiltonian: 
\begin{eqnarray}
H_{K} &=&\frac{1}{2}\sum\limits_{k,k^{\prime },\alpha ,\beta }(J^{z\uparrow
}c_{k\alpha \uparrow }^{\dag }c_{k\beta \uparrow }-J^{z\downarrow
}c_{k\alpha \downarrow }^{\dag }c_{k\beta \downarrow })S^{z}  \nonumber \\
&+&J^{\perp }(c_{k\alpha \downarrow }^{\dag }c_{k\beta \uparrow
}S^{+}+c_{k\alpha \uparrow }^{\dag }c_{k\beta \downarrow }S^{-}).
\end{eqnarray}
The bare values of the Kondo couplings are 
\beq
J^{\perp }=J^{z\uparrow}=J^{z\downarrow }\sim 2t^{2}
\left(\frac{U}{(|\varepsilon _{d}|(\varepsilon _{d}+U)}\right)=J,
\eeq
where $\varepsilon _{d}\sim \varepsilon_{d\uparrow }\sim \varepsilon
_{d\downarrow }$. Note also that $\psi _{L\uparrow }$ is a linear
superposition of $\psi _{L+}$ and $\psi _{L-}$. One can then perform the
second stage of the RG analysis directly on the Kondo Hamiltonian. 
The RG equations can be obtained by
integrating out the high energy degrees of freedom between a cut-off $
\Lambda _{0}\sim {\varepsilon}_{d}$ and $\Lambda $ and read: 
\begin{eqnarray}
\frac{d\lambda _{z\uparrow }}{d\ln l} &=&2\left[ \cos ^{2}(\theta /2)+\frac{
1+p}{1-p}\sin ^{2}(\theta /2)\right] \lambda _{\perp }^{2}, \\
\frac{d\lambda _{z\downarrow }}{d\ln l} &=&2\left[ \cos ^{2}(\theta /2)+
\frac{1-p}{1+p}\sin ^{2}(\theta /2)\right] \lambda _{\perp }^{2}, \\
\frac{d\lambda _{\perp }}{d\ln l} &=&\left[ \cos ^{2}(\theta /2)+\frac{1-p}{
1+p}\sin ^{2}(\theta /2)\right] \lambda _{\perp }\lambda _{z\uparrow } 
\nonumber \\
&+&\left[ \cos ^{2}(\theta /2)+\frac{1+p}{1-p}\sin ^{2}(\theta /2)\right]
\lambda _{\perp }\lambda _{z\downarrow },
\end{eqnarray}
where we have introduced the dimensionless Kondo couplings $\lambda _{\perp }=
\sqrt{\rho _{+}\rho _{-}}J^{\perp }$, $\lambda _{z\uparrow /\downarrow
}=\rho _{\pm }J^{z,\uparrow /\downarrow }$ and $l=\Lambda _{0}/\Lambda $.
The dimensionless Kondo couplings are all driven to strong coupling at the
same energy scale $T_{K}$ which can be determined explicitly by integrating
out the above RG equations. The Kondo temperature can be expressed as: 
\begin{equation}
T_{K}(p,\theta )=\Lambda _{0}\exp \left( -\frac{\mathrm{arctanh}(p\cos
(\theta ))}{2\lambda _{0}p\cos (\theta )}\right) ,  \label{tk}
\end{equation}
where $\lambda _{0}=J_{0}(\rho _{+}+\rho _{-})$. This expression generalizes
the one obtained in Ref. [\onlinecite{martinek1}] to all values of $\theta $. 
The
dependence of the Kondo temperature with $p,\theta $ appears as a function
of the single parameter $p\cos (\theta )$ and is therefore a uniform
function of $\theta $. $T_{K}$ takes its maximum value $T_{K}^{0}$
independent of $p$ at $\theta =\pi /2$ which corresponds to antiparallel
magnetizations and $T_{K}$ is minimum for $\theta =0$
as expected.

\section{Numerical renormalization group analysis} \label{sec:nrg}

In order to compute dot spectral
densities and transport properties we proceed with a NRG analysis.
At this point it is convenient to 
perform two unitary transformations on the initial
Hamiltonian defined in Eq. (\ref{eq:h}). We first make an even/odd 
transformation introducing the operators
\beq 
c_{k,e/o,\pm}=(c_{k,L,\pm}\pm c_{k,R,\pm})/\sqrt{2}.
\eeq
We next introduce a new basis defined by
\beq
b_{k,e,\uparrow/\downarrow}=c_{k,e,\pm}~~,~~b_{k,o,\uparrow/\downarrow}=c_{k,e,\mp}.
\eeq
The tunneling Hamiltonian can be written in the new basis as: 
\begin{equation}
H_{tun}=\sum\limits_{k,\eta=e/o,\sigma}(t_\eta
b^\dag_{k\eta,\sigma} d_\s+H.c.)
\end{equation}
where $t_e=t\sqrt{2}\cos(\theta/2)$, and $t_o=t\sqrt{2}\sin(\theta/2)$.
The crucial point is that the angle $\tht$ arising due to the non collinear lead magnetizations 
is now hidden in the tunneling amplitudes $t_e$ and $t_o$.

Note that in this formulation the dot is coupled with a single effective lead with an
angle and spin dependent bulk spin DOS given by
\beq
\rho^{eff}_{\uparrow/\downarrow}=\rho_{\pm}\cos^2(\theta/2)+\rho_{
\mp}\sin^2(\theta/2),
\eeq
which is mixture of the initial bulk spin densities of
states. 
This allows us to treat this problem in a simplified way using
the numerical renormalization group method \cite{hewson-book,hofstetter,costi}
with the generalizations to treat arbitrarily shaped densities of states. \cite{ingersent}

We insist that such formulation has been made possible only because
 we consider symmetric tunneling amplitudes $t_L=t_R$. Note that for 
antiparallel polarizations ($\theta=\pi/2$) we have $\rho^{eff}_{\uparrow}=\rho^{eff}_{\downarrow}$
and the effective model has spin symmetry.

\subsection{Study of the dot density of states}

From hereon we will consider the DOS along the polarization
direction in each lead to be given by the tight-binding expressions
\begin{equation}
\rho _{\pm }(\omega )=\frac{2}{\pi D}\sqrt{1-\left( \frac{\omega +\delta
\epsilon _{\pm }}{D}\right) ^{2}},  \label{eq:freedens}
\end{equation}
where $\delta \epsilon _{+}=-\delta \epsilon _{-}=\delta \epsilon $ is the
spin-dependent shift of the bands with respect to the Fermi level. The
quantity $\delta \epsilon $ characterizes the  magnetization $P$ of the
leads given by 
\begin{equation}
P=\int_{-\infty }^{\infty }d\omega f(\omega )[\rho _{+}(\omega )-\rho
_{-}(\omega )],  \label{eq:polariz}
\end{equation}
where $f(\omega )$ is the Fermi function. Note that the densities of states in the leads
are no longer constant which is more realistic to describe ferromagnetic leads. \cite{konig04}
For the sake of simplicity, from
hereon we take the chemical potential equal to zero. Note that for this
particular choice the parameter $p$ defined in the previous section is zero
while the spin polarization is not.  For small $\delta \epsilon $ and
temperatures the magnetization is simply given by $P\simeq \frac{4\delta
\epsilon }{\pi D}$. In the present case, the most important effect of the
leads magnetism is the occurrence of an internal field $B_{int}$ that creates
the effective Zeeman splitting already mentioned above. 

To calculate the dot density of states we need the local Green function. In the standard form, we have
\beq
G^{-1}_\sigma{(\omega{})}=G^{-1}_{0\sigma}(\omega)-\Sigma(\omega),
\eeq
where $\Sigma_\sigma(\omega)$ is the self-energy and $G_{0\sigma}(\omega)^{-1}$ is the non-interacting Green function
\beq
G_{0\sigma}(\omega)=\frac{1}{\omega-\epsilon_{d,\sigma}-g\mu B \sigma-\Delta_\sigma(\omega)}.
\eeq
The hybridization function $\Delta_{\sigma }(\omega )$
is spin-dependent and has in general a non-zero real part that is given by 
\begin{equation}
{\cal R}[\Delta_{\sigma }(0)]=\sigma 4t^{2}\delta \epsilon \cos (\theta
)/D^{2},  \label{eq:realhybr}
\end{equation}
for our choice of the DOS and $\omega\sim 0$.
This is equivalent to an angle-dependent effective
magnetic field 
\beq\mu _{B}B_{int}=8t^{2}\delta \epsilon \cos (\theta )/D^{2},\eeq
that polarizes the dot's spin and  destroys the Kondo effect for $\mu
_{B}B_{int}\gtrsim T_{K}$. 

The first thing we want to check is whether the polarized leads with a given
angle $\theta$ are able to split the dot DOS [$\rho_\sigma(\omega) =-{\cal I}[G_\sigma(\omega)]/\pi$]. It has been
shown that polarized leads with antiparallel polarized directions
(corresponding to $\theta =\pi /2$ here) do not split the dot density of
states.\cite{lopez1,martinek1} Is there a finite interval around $\theta
=\pi /2$ where this property holds? We have plotted in Fig. 1
the dot
spectral density $\rho _{\ua}$ in the spin up sector for different values of 
$\theta $. As expected the low energy peak of $\rho _{\ua}$ is pinned at $\omega =0$
for $\theta =0$ but is shifted away from $\omega =0$ as soon as $\theta $ is
decreased. This is much clearer in the right inset of Fig. 1
where the peak is zoomed.

\begin{figure}[h]
\label{Fig:DOS1} \includegraphics[width=8.5cm,clip=true]{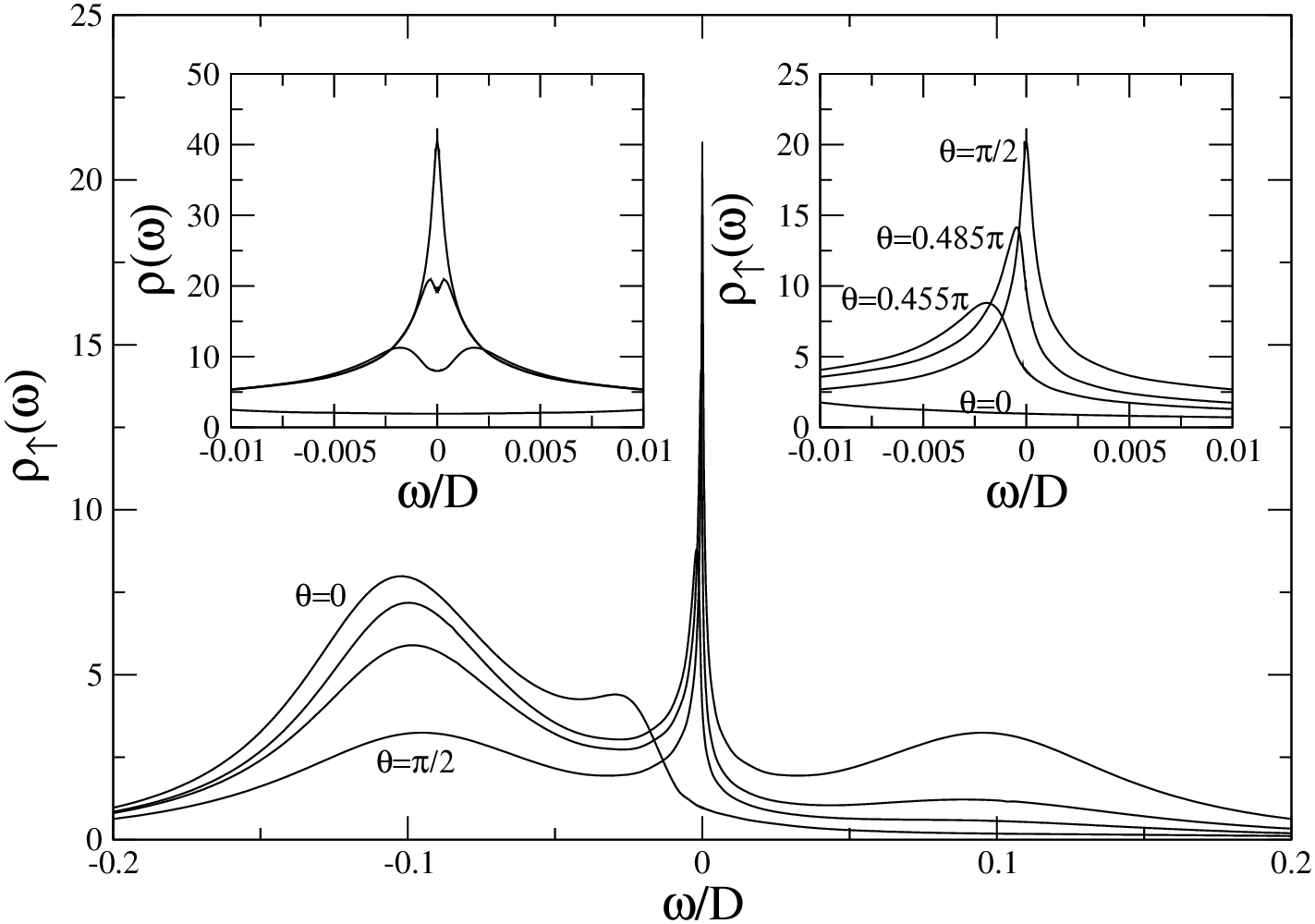} 
\caption{Dot density of states for the spin-up sector as function of
energy for different values of $\theta=0,\,0.455\pi,\,0.485\pi,\,\pi/2$.
The shift of the bands is fixed to $\delta\epsilon=0.5D$. Other parameter are: $U=0.2$, 
$\epsilon{_d=-U/2}$, and $t=0.071$. Right inset: zoom of $\rho_\ua$ for the same values of $\theta$. We
clearly see the shift of the maximum $\rho_\ua$ when decreasing $\theta$ away
from $\theta=\pi/2$. Left inset: Total density of states as function of
frequency for the four aforementioned values of $\theta$. The density of
states is split as soon as the effective magnetic field is larger than the Kondo
temperature.}
\end{figure}

The splitting of the total dot density of states $\rho=\rho_\ua+\rho_\da$ is
shown in the left inset of Fig. 1 for four different values of $\theta$. The
splitting gradually increase as $\theta$ is decreased from $\theta=\pi/2$
and is maximum for $\theta=0$. {\pbf For very small deviations of
  $\tht$ around $\pi/2$, the effective magnetic field $\mu_BB_{int}
  \lesssim T_K$ and no splitting is expected.}

\begin{figure}[h]
\includegraphics[width=8.5cm,clip=true]{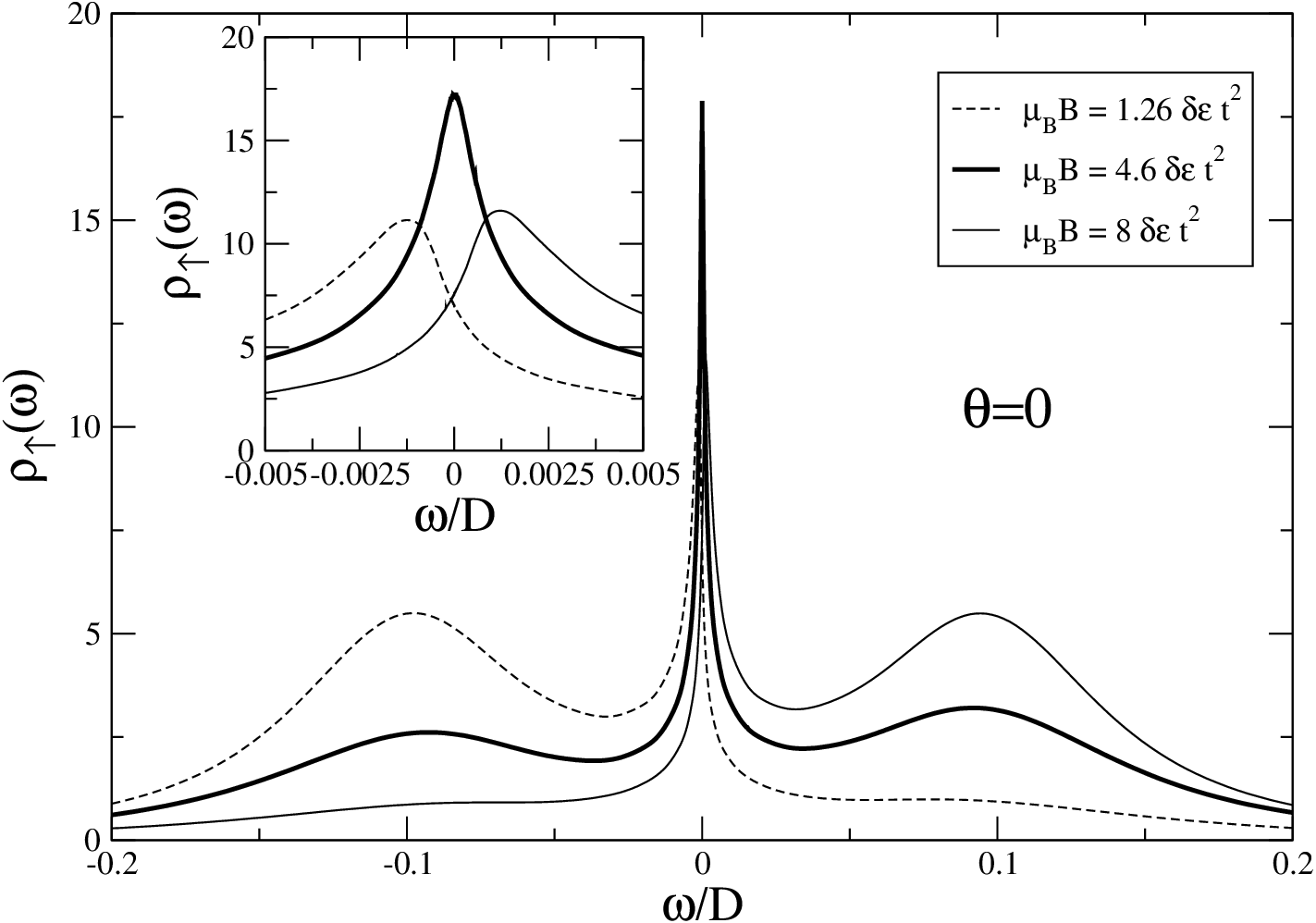}
\caption{Dot density of states for the spin-up sector as function of
energy for different values of the magnetic field in the case with
parallel magnetizations ($\theta =0$).  Other parameters as in Fig.~1.
The spin polarization in the dot and the
splitting of the Kondo peak can be compensated with a magnetic field.}\label{Fig:DOS2} 
\end{figure}

\begin{figure}[h]
\includegraphics[width=8.5cm,clip=true]{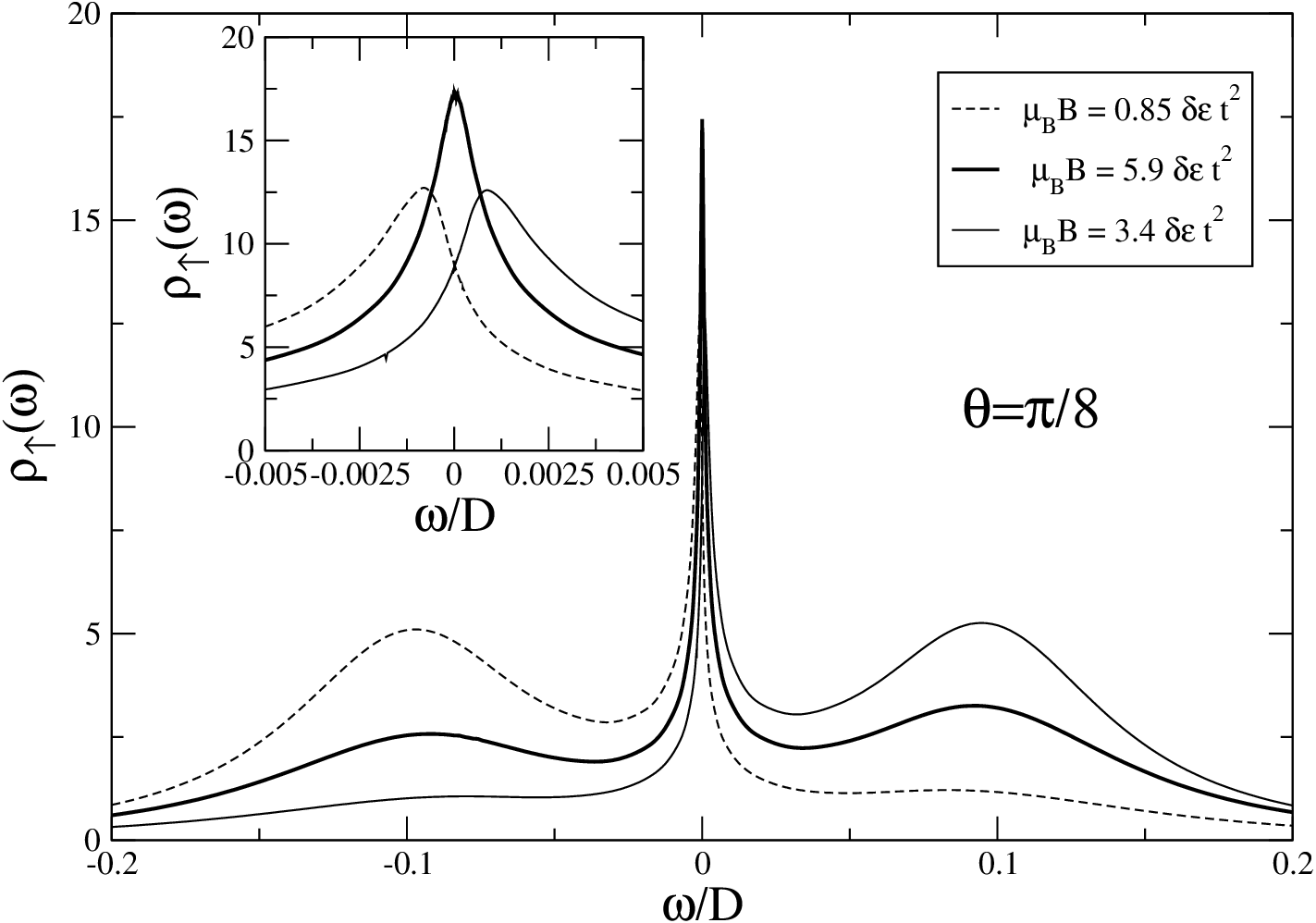}
\caption{Same as Fig. \ref{Fig:DOS2} with $\theta=\pi/8$.
}\label{Fig:DOS3} 
\end{figure}

For collinear polarization directions, it has been shown that this splitting
can be compensated by an external magnetic field \cite{martinek2} restoring
the Kondo effect. {\pbf For general directions of the lead polarizations, 
an effective magnetic field is also generated. This effective magnetic field can be compensated
by applying an external magnetic field in the plane $(\vec n_L,\vec n_R)$ with a direction opposite
to $\vec e_z$. We therefore expect that the splitting of the Kondo zero bias anomaly induced
by the polarized electrodes can be corrected by an adequate magnetic field for all values of $\theta$. We have checked that this property holds using NRG. 
We have presented in Fig. \ref{Fig:DOS2} the DOS $\rho _{\ua}$ as function of energy for
$\theta =0$ (where the splitting is maximum) and for different value of the external magnetic
field. For a given value of the external magnetic field, the splitting of the Kondo peak 
vanishes and the Kondo peak is restored at $\omega=0$. This is also true for a general angle $\theta$ (see Fig.~\ref{Fig:DOS3}).
These properties of the dot's spectral density will clearly reflect  in the transport
properties of the system as we will see.}

\subsection{Transport properties} \label{sec:transport}
\begin{figure}[h]
\includegraphics[width=8.5cm,clip=true]{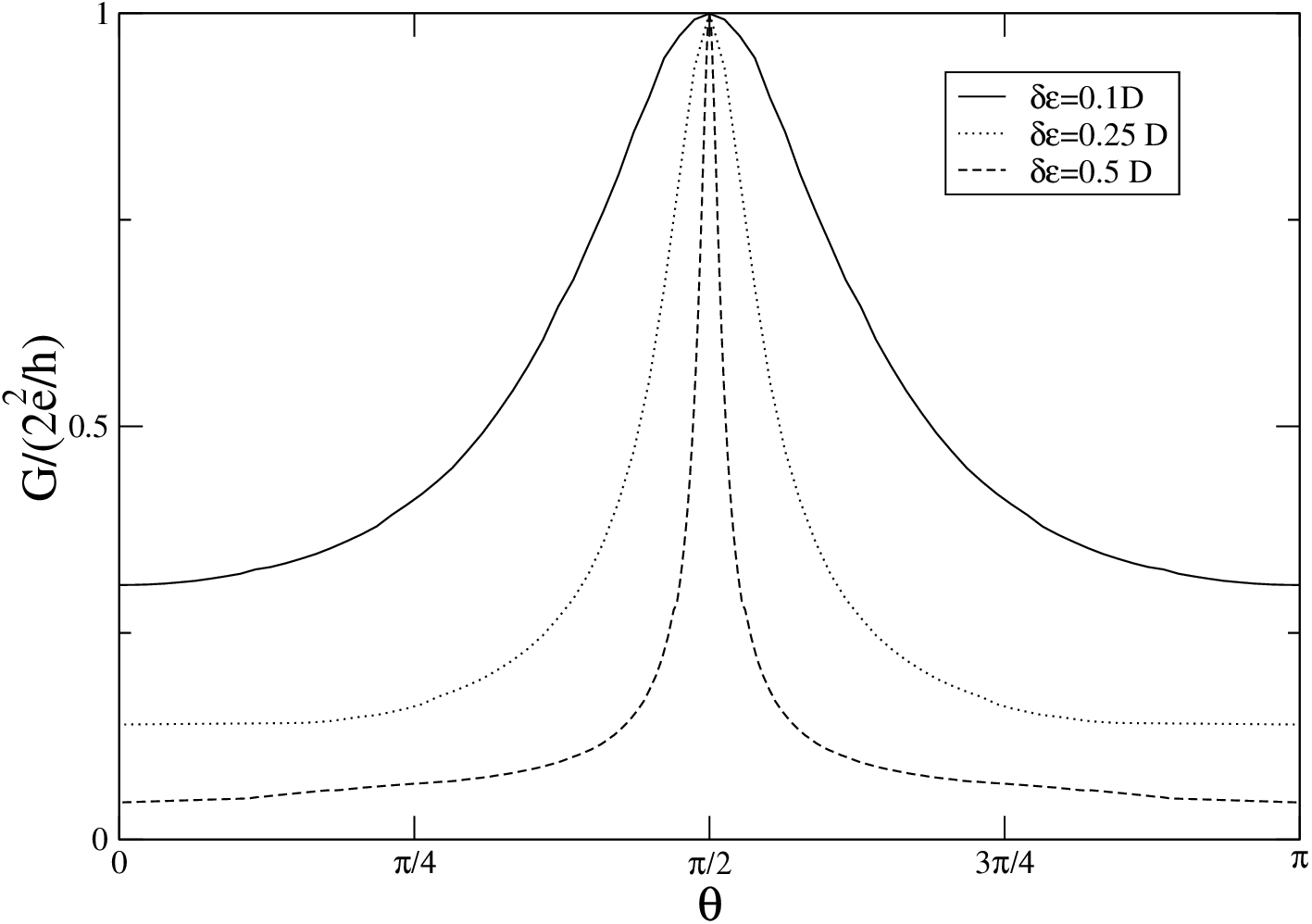}
\caption{Conductance as a function of the angle for different values of the
bands shift $\delta\epsilon=0.1D,\,0.25D,\,0.5D$. Other parameters as in Fig.~1.}\label{Fig:conductance}
\end{figure}
Using the Keldysh formalism, it has been shown
that for the collinear magnetizations, as in the case of non-magnetic leads,
the finite temperature conductance can be put in terms on the dot's spectral
density. \cite{mw} This is not the case for arbitrary angles. The problem is that for
non collinear magnetizations, the two spin channel of the $L$ and $R$ leads
are mixed and the standard approach used to eliminate the lesser propagator
can not be applied. To our knowledge, in these circumstances, the only
direct way of evaluating the conductance at any temperature is through the
Kubo formalism. This requires evaluating two particle propagators, a matter
that requires sophisticated numerical codes. There are however some limiting
cases that illustrate the general behavior as we will see. 
{\pbf In our NRG formulation we have $\rho_{d\ua}(0)=\rho_{d\da}(0)$.
Since the DOS in the leads are energy dependent (see Eq. (\ref{eq:freedens})), 
this equality does not prevent to
have a non zero polarization following Eq. (\ref{eq:polariz}) though $p=0$. 
For this case, the hybridization matrices $\Gamma_{\s\s'}=-{\cal I}(\Sigma_{\s\s'})$ 
(where $\s,\s'=\ua,\da$) are diagonal and {\it angle independent}.
This  allows the use of the Meir Wingreen formula. \cite{mw}
The 
$T=0$ conductance is thus simply proportional to $|G_{d}(0)|^2$.
The $T=0$ conductance has been computed using NRG and is shown in
Fig. \ref{Fig:conductance} as a function of $\tht$ 
for different values of the splitting $\delta \veps$. The conductance reaches the unitary limit
 for $\tht=\pi/2$ (antiparallel case) and is minimum for parallel magnetizations.
Though this result may look unusual, this is expected since the splitting is zero for $\tht=\pi/2$ restoring the Kondo effect
and is maximum for $\tht=0$. Remember that $p(0)=0$ with the
definition of the DOS we used in NRG.}
 
{\pbf This result should be contrasted to the situation where 
an external magnetic field able to restore the Kondo zero bias
anomaly is added.
We consider constant bulk DOS 
as in Sec. \ref{sec:rg}.  Electrons
with spins $\ua$ have a different DOS at the Fermi energy than
electrons with spins $\da$, resulting in $p(0)\ne 0$ (as in [\onlinecite{lopez1,martinek1,martinek2}]).
At $T=0$, the Kondo singlet is therefore formed and the Fermi liquid theory
can be applied. The 
conductance $G$ through the device reaches at $T=0$ its
maximum conductance $G_{u}$ which reads: 
\begin{equation}
G_{u}=\frac{2e^{2}}{h}\frac{1-p^{2}}{1-p^{2}\cos^{2}(\theta )}~.
\end{equation}
Notice that here $G$ reaches the unitary limit for $\tht=0$ whatever the
value of $p$ and is minimum for $\tht=\pi/2$ at fixed $p$.
In order to determine to determine $G_u$, we have extended the scattering approach developed by 
Ng and Lee \cite{ng} to polarized electrodes following [\onlinecite{martinek2}].
This anomalous denominator comes from the fact the imaginary part of the dot retarded Green function
$G_{d\s}^r$ is simply $-1/\Gamma_{\s\s}$ at $T=0$ which depends on the angle $\tht$ and $p$.}
At finite $T\ll T_{K}(p,\theta )$, deviations of order $(T/T_K)^2$ from $G_{u}$ are
expected using the Fermi liquid theory. 
On the other hand, at high
temperature $T\gg T_{K}(p,\theta )$, the conductance is much smaller and can
be computed by renormalized perturbation theory: 
\begin{equation}
G=\frac{2e^{2}}{h}(1+p^{2}\cos (2\theta ))\frac{3\pi ^{2}/16}{\ln
^{2}(T/T_{K})}.
\end{equation}
Notice that the expression for the high temperature conductance has the same 
angular expression as the one for a tunnel junction between two ferromagnetic leads 
conversely to the low temperature case.

\section{Conclusion} \label{sec:conclusion}
In this paper, we have studied a quantum dot in the Kondo regime
between two non collinear ferromagnetic electrodes by combining poor man scaling and NRG
techniques. We have shown using NRG that the Kondo zero bias anomaly is in general splitted
for all angles between the two lead magnetizations (except for the antiparallel case).
This splitting can be compensated by an external magnetic field. Based on the spectral
densities, we have addressed transport properties for both compensated and uncompensated cases.

{\it Note added} Some of the results presented in this paper have been also independently
obtained in a recent preprint \cite{eto}. \vskip 0.2cm 
\textit{
Acknowledgments.} -- PS would like to thank interesting communications with
J. Martinek. Part of this research was supported by the contract PNANO
`QuSpins' of the French Agence Nationale de la Recherche.

\end{document}